\documentclass[a4paper,12pt]{article}
\usepackage{mathrsfs}
\usepackage{epsfig}
\usepackage[dvips,usenames]{color}
\usepackage{graphicx}

\newlength{\dinwidth}
\newlength{\dinmargin}
\let\jnfont=\rm
\def\NPB#1,{{\jnfont Nucl.\ Phys.\ B }{\bf #1},}
\def\PLB#1,{{\jnfont Phys.\ Lett.\ B }{\bf #1},}
\def\EPJC#1,{{\jnfont Eur.\ Phys.\ Jour.\ C }{\bf #1},}
\def\PRD#1,{{\jnfont Phys.\ Rev.\ D }{\bf #1},}
\def\PRL#1,{{\jnfont Phys.\ Rev.\ Lett.\ }{\bf #1},}
\def\MPLA#1,{{\jnfont Mod.\ Phys.\ Lett.\ A }{\bf #1},}
\def\JPG#1,{{\jnfont J.\ Phys.\ G}{\bf #1},}
\def\CTP#1,{{\jnfont Commun.\ Theor.\ Phys.\ }{\bf #1},}
\begin{document}

\begin{center}
{\Large Effects of R-parity violating supersymmetry in top pair production

        at linear colliders with polarized beams }
\vspace{.2in}

         Xuelei Wang, Jitao Li, Suzhen Liu
\vspace{.2in}

{\it College of Physics and Information Engineering, Henan Normal University,\\
                                                     Xinxiang  453007, China}

\end{center}

\begin{abstract}

In the minimal supersymmetric standard model with R-parity
violation, the lepton number violating top quark interactions can
contribute to the top pair production at a linear collider via
tree-level \emph{u}-channel squark exchange diagrams. We calculate
such contributions and find that in the allowed range of these
R-violating couplings, the top pair production rate as well as the
top quark polarization and the forward-backward asymmetry can be
significantly altered. By comparing the unpolarized beams with the
polarized beams, we find that the polarized beams are more powerful
in probing such new physics.

\end{abstract}

\newpage
\begin{center}
{\bf I. Introduction}~~
\end{center}

The top quark has an exceptionally large mass of the order of the
electroweak symmetry breaking scale and is naturally expected to
have a close connection to new physics \cite{1a,1b}. The most
popular model of new physics is the Minimal Supersymmetric Model
(MSSM). In this model, a discrete multiplicative symmetry of
\emph{R}-parity, defined by $\emph{R}=(-1)^{2S+3B+L}$, with spin
\emph{S}, baryon number \emph{B} and lepton number \emph{L}, is
often imposed on the Lagrangian to maintain the separate
conservation of \emph{B} and \emph{L}. However this conservation
requirement is not dictated by any fundamental principle such as
gauge invariance or renormalizability. The finiteness of the
neutrino mass as suggested by the Super-Kamiokande and several other
neutrino experiments also implies that lepton numbers may be
violated. So it is of great interest to consider possible violation
of this symmetry and study its experimental consequences in collider
experiments.

The most general superpotential of the MSSM consistent with the
$SU(3)\times SU(2)\times U(1)$ symmetry and supersymmetry contains
\emph{R}-violating interactions which are given by \cite{5}
\begin{eqnarray}
 \emph{W}_{R\!\!\!/}=\frac{1}{2}\lambda_{ijk}L_{i}L_{j}E^{c}_{k}
 +\lambda^{'}_{ijk}\delta^{\alpha\beta}L_{i}Q_{j\alpha}D^{c}_{k\beta}
 +\frac{1}{2}\lambda^{''}_{ijk}
\epsilon^{\alpha\beta\gamma}U_{i\alpha}^{c}D_{j\beta}^{c}D^{c}_{k\gamma}
 +\mu_{i}L_{i}H_{2}
\end{eqnarray}
Here $L_{i} (Q_{i})$ and $E_{i} (U_{i},D_{i})$ are the left-handed
lepton (quark) doublet and right-handed lepton (quark) singlet
chiral superfields, and $\emph{c}$ denotes charge conjugation.
$H_{1,2}$ are the Higgs chiral superfields. The indices  $\emph{i}$,
$\emph{j}$, $\emph{k}$  denote generations and $\alpha$, $\beta$ and
$\gamma$ are the color indices. We focus our attention only on the
tri-linear supersymmetric \emph{R}-parity violating interactions in
Eq. (1) and assume that the bi-linear terms $\mu_{i}L_{i}H_{2}$ can
be rotated away by a field redefinition \cite{6}. The $\lambda$ and
$\lambda^{'}$ are the coupling constants of the $\emph{L}$-violating
interactions and $\lambda^{''}$ those of the $\emph{B}$-violating
interactions. The non-observation of the proton decay imposes very
strong constraints on the products of the $\emph{L}$-violating and
$\emph{B}$-violating couplings \cite{7}. It is thus conventionally
assumed in phenomenological studies that only one type of these
interactions (either $\emph{L}$- or $\emph{B}$-violating) exists.
Constraints on these $\emph{R}$-parity violating couplings have been
obtained from various low-energy process and it is notable that
the bounds on the couplings involving top quark are generally quite
weak (see \cite{25} for a review). In the future precision top
quark experiments, these couplings may either manifest themselves or
subject to further constraints.

The \emph{R}-parity violating interactions can have rich
phenomenology in top quark physics: they can have sizable effects in
top quark decays and productions at colliders. For example, the
\emph{R}-violating couplings can greatly enhance the FCNC top quark
decays \cite{22} and can induce new mechanisms for the productions
of single top \cite{21} and top pair at hadronic colliders
\cite{29}. At a linear collider, the \emph{R}-violating couplings
can contribute to the top pair production via tree-level
\emph{u}-channel squark exchange diagrams. In \cite{18} the author
has investigated such contributions to the production rate for
unpolarized beams. Considering that the future linear collider will
likely have polarized beams, we extend the study to the case of
polarized beams. By comparing the polarized beams with the
unpolarized beams, we find that the polarized beams are more
powerful in probing such \emph{R}-violating interactions. Also, we
will study the effects of these \emph{R}-violating couplings in top
quark polarization as well as the forward-backward asymmetry at a
linear collider.

This article is arranged as follows. In Sec. II we will calculate
the effects of the \emph{R}-violating couplings in top pair
production at a linear collider.  In Sec. III we will present the
numerical results and give some discussions. And finally in Sec. IV
we give the conclusion.

\begin{center}
{\bf II. Calculations}
\end{center}

In terms of four-component Dirac spinors, the second term in Eq.(1),
$\lambda^{'}_{ijk}\delta^{\alpha\beta}L_{i}Q_{j\alpha}D^{c}_{k\beta}$,
leads to the following inetractions
\begin{eqnarray}
\mathcal{L}_{\lambda^{'}}=-\lambda^{'}_{13k}
 (\widetilde{d}^{k}_{R})^{*}(\overline{e}_{L})^ct_{L}+h.c.
\end{eqnarray}
Apart from the s-channel diagrams mediated by $\gamma$ or $Z$, the top
pair production can also proceed through the diagram mediated by the
squark shown in Fig.1 due to such lepton-number violating interactions.
Looking at the lagrangians, one can easily find that in \emph{R}-parity
violating inetractions, one vertex is proportional to $\emph{P}_{L}$ and the
other is proportional to $\emph{P}_{R}$ ($\emph{P}_{R,L}=\frac{1}{2}(1\pm\gamma_{5}$)).
\begin{figure}[ht]
\begin{center}
\epsfig{file=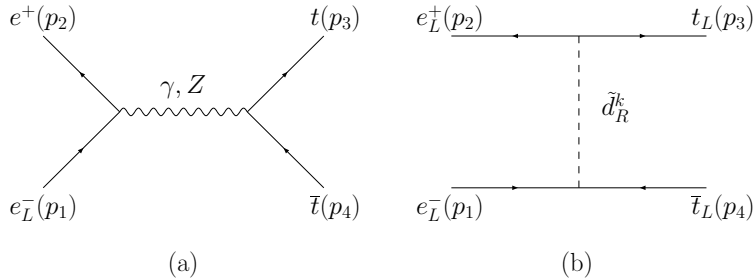,width=10cm}
\caption{\small Feynman diagrams for $e^+e^-_L\rightarrow t\overline{t}$: (a) the SM diagrams,
         (b) the \emph{R}-violating diagrams.} \label{fig1}
\end{center}
\end{figure}
The expressions of the amplitudes, together with the SM contributions \cite{18}, are given
by
\begin{eqnarray}
\mathcal{M}_{SM}&=&-\frac{1}{s-m_{V}^{2}+\emph{i}\emph{m}_{V}\Gamma_{V}}
                 (\bar{\emph{u}}(\emph{p}_{3})
\gamma^{\mu}(\emph{a}_{t}+\emph{b}_{t}\gamma_{5})\emph{v}(\emph{p}_{4}))
(\bar{\emph{v}}(\emph{p}_{2})\gamma_{\mu}(\emph{a}_{e}
+\emph{b}_{e}\gamma_{5})\emph{u}(\emph{p}_{1}))\\
\mathcal{M}^{'}_{SM}&=&-\frac{1}{s-m_{V}^{2}
+\emph{i}\emph{m}_{V}\Gamma_{V}}(\bar{\emph{u}}(\emph{p}_{3})
\gamma^{\mu}(\emph{a}_{t}+\emph{b}_{t}\gamma_{5})\emph{v}(\emph{p}_{4}))
(\bar{\emph{v}}(\emph{p}_{2})\gamma_{\mu}(\emph{a}_{e}
+\emph{b}_{e}\gamma_{5})P_{L}\emph{u}(\emph{p}_{1}))\\
\mathcal{M}_{\lambda^{'}}&=&\frac{\mid\lambda^{'}\mid^2}{u-m_{\tilde{d}}^{2}}
(\bar{u}(\emph{p}_{3})P_{R}v^{c}
(\emph{p}_{2}))(\bar{u}^{c}(\emph{p}_{1})P_{L}v(\emph{p}_{4}))
\end{eqnarray}
Here $\mathcal{M}_{SM}$ stands for the two SM s-channel diagrams mediated by $\gamma$ and
$Z$ for unpolarized beams.
For the photon-exchange diagram, $\emph{a}_{t}=\frac{2}{3}e, \emph{a}_{e}=-e,
\emph{b}_{t}=\emph{b}_{e}=0$ and $\emph{m}_{V}=\Gamma_{V}=0$. For
the Z-exchange diagram, $\emph{a}_{t}=\frac{g}{4\cos\theta_{W}}(1-\frac{8}{3}\sin^{2}\theta_{W}),
\emph{b}_{t}=-\frac{g}{4\cos\theta_{W}},
\emph{a}_{e}=\frac{g}{4\cos\theta_{W}}(4\sin^{2}\theta_{W}-1),
\emph{b}_{e}=\frac{g}{4\cos\theta_{W}}$.
$\mathcal{M}^{'}_{SM}$ is same as $\mathcal{M}_{SM}$, but for left-handed
electron beam. $\mathcal{M}_{\lambda^{'}}$ is for squark-mediated diagram.
The Mandelstum variables are defined as $s=(p_{1}+p_{2})^2$,  $t=(p_{1}-p_{3})^2$
and $u=(p_{1}-p_{4})^2$. The amplitude for squark-mediated diagram is proportional
to $|\lambda^{'}|^2$.

Note that the \emph{R}-parity violating couplings involved in our calculations
are $\lambda^{'}_{13k}$ with $k=1,2,3$ being the generation index.
We will consider only one such coupling at a time. From \cite{25} we know
that the coupling $\lambda^{'}_{132}$ is the most weakly constrained one.
So we will use this coupling in the following analyses.
This implies that the exchanged squark in Fig.1(b) is the supersymmetric
partner of s-quark.

\begin{center}
{\bf III. Numerical results and discussions}
\end{center}

In this section we present  some numerical results. In our
calculations we only consider the tree-level diagrams, and for the
\emph{R}-violating contributions we only consider its interference
with the corresponding SM diagrams and neglect the higher order
contributions. The current constraints on $\lambda^{'}_{132}$ scales
linearly with $\widetilde{t}_{L}$ mass, which is smaller than 0.28
for $\widetilde{t}_{L}= 100$ GeV \cite{25}. To be conservative we
will take $\lambda^{'}_{132}=0.1$ for illustration. For other
parameters involved, we take $m_{t}=175$ GeV, $m_{Z}=91$ GeV,
$\alpha=1/128$ and $\sin^{2}\theta_{w}=0.23$. Note that around the
center-of-mass energy of 350 GeV $(\sim 2 m_{t})$, the threshold
effects are very important \cite{27}. In our calculations we avoid
this by considering a center-of-mass energy above the
$t\overline{t}$ threshold.
\begin{figure}[ht]
\begin{tabular}{cc}
\epsfig{file=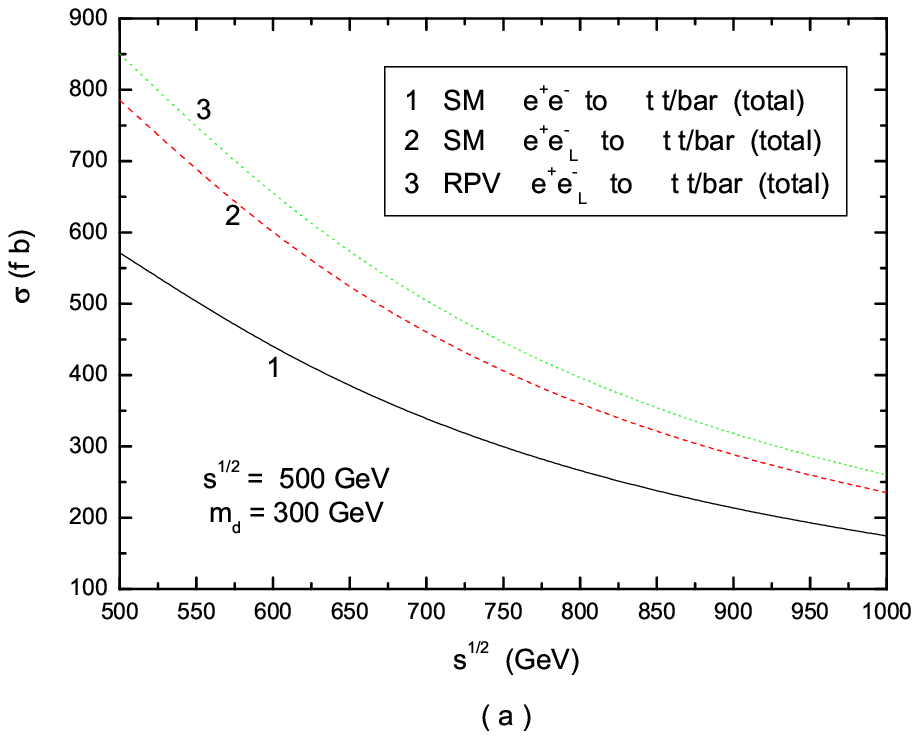,width=235pt,height=235pt}\\
\end{tabular}
\begin{tabular}{cc}
\epsfig{file=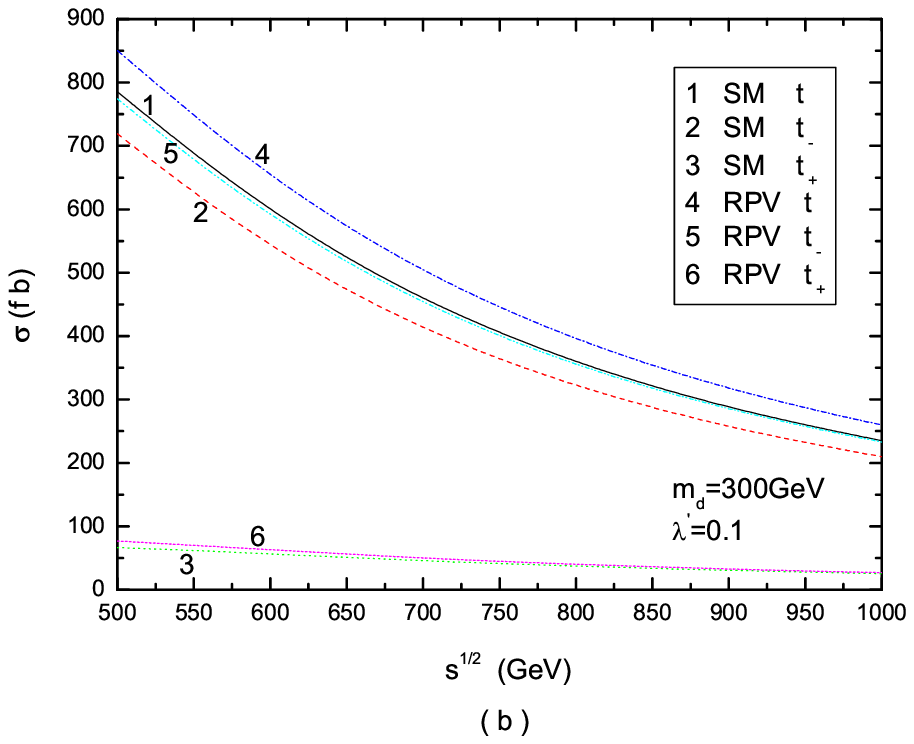,width=235pt,height=235pt}\\
\end{tabular}
\vspace*{-1.1cm}
\caption{\small The cross section of $e^+e^-\rightarrow t\overline{t}$
         versus the center-of-mass energy.}
\label{fig2}
\end{figure}
In Fig. 2 we plotted the cross section of top pair production versus
the center-of-mass energy in various cases. For the purpose of
illustration, we present the cross section with squark mass of 300
GeV, which is well above the bounds given by CDF and D0 \cite{28}.
In Fig.2(a) the curves are for the predictions in the SM and
\emph{R}-violating SUSY with unpolarized beams and the left-handed
polarized $e^{-}$ beam, respectively. In Fig.2(b), we used the
left-handed polarized $e^{-}$ beams and showed the cross section of
$t\bar t$ with definite helicity of the top quark, where $t_+$
($t_-$) stands for a top quark with positive (negative) helicity. It
is clear from Fig.2(a) that the left-handed polarized $e^{-}$ beams
lead to significant enhancement for the production cross section.
From both figures we can see that the presence of lepton number
violating interactions can increase the cross section by over 10
percent.

Now we consider the dependence of the cross section on squark
mass. For this purpose, we fix the center-of-mass energy at 500 GeV.
In Fig.3(a) we plot the cross-section versus the squark mass. As
the squark mass increases, the cross-sections converge to their
corresponding SM values, indicating the decoupling nature of the
squark interactions. In Fig.3(b), we plot the cross section
versus the coupling constant $\lambda^{'}$.
We can see that as $\lambda^{'}$ increases, the cross sections become
larger than their corresponding SM values since
the cross section is proportional to $|\lambda^{'}|^2$.
\begin{figure}[ht]
\begin{tabular}{cc}
\epsfig{file=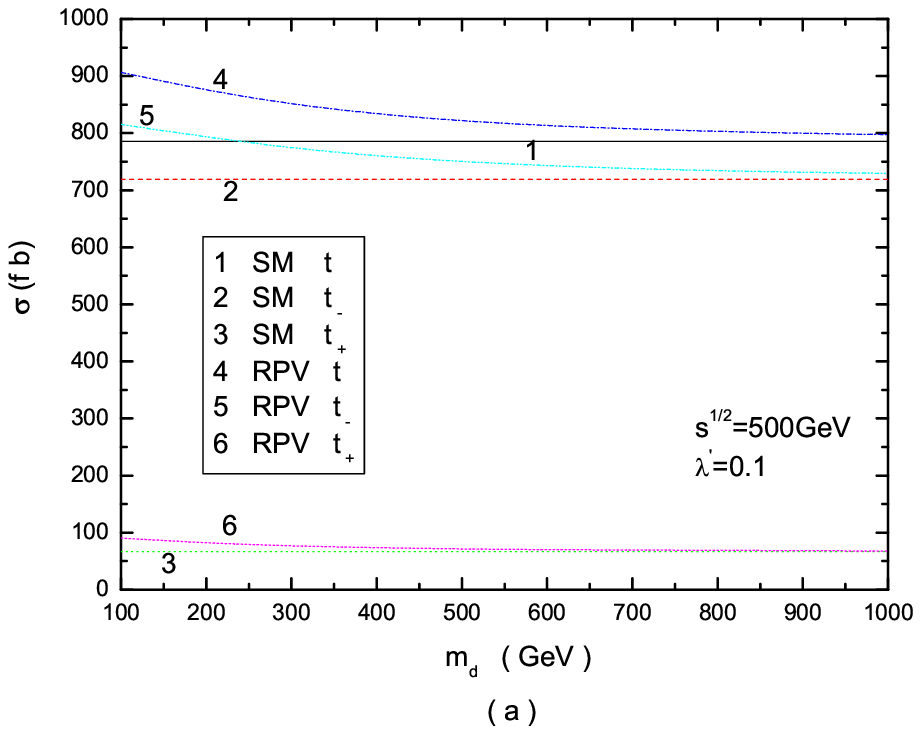,width=225pt,height=235pt}\\
\end{tabular}
\begin{tabular}{cc}
\epsfig{file=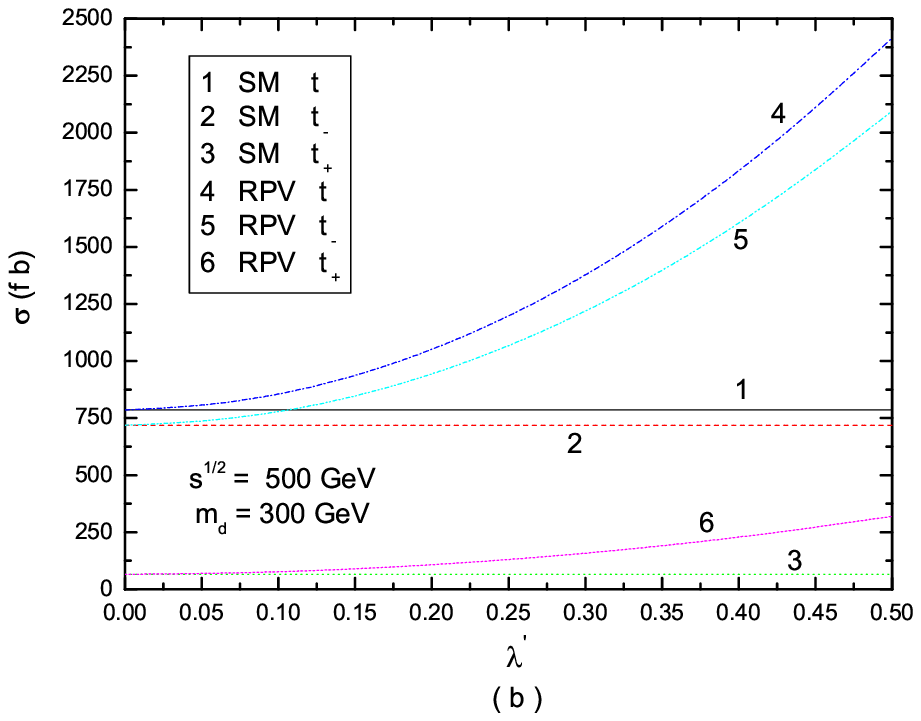,width=225pt,height=235pt}\\
\end{tabular}
\vspace*{-0.8cm} \caption{\small The cross section of
$e^+e_{L}^-\rightarrow t\overline{t}$ versus (a) the intermediate
squark mass and (b) the \emph{R}-violating coupling $\lambda^{'}$.}
\label{fig3}
\end{figure}

In addition to the production rate, we also present the effects of
the \emph{R}-violating couplings on the top polarization defined by
\begin{eqnarray}
\emph{P}_{t}=\frac{\widehat{\sigma}_{-}-\widehat{\sigma}_{+}}
             {\widehat{\sigma}_{-}+\widehat{\sigma}_{+}},
\end{eqnarray}
where $\widehat{\sigma}_{-}$ ($\widehat{\sigma}_{+}$) indicates the
cross section of $t\bar{t}$ with negative (positive) helicity of
the top quark. The hemisphere top polarizations are obtained by
restricting the top scattering angle in the $t\overline{t}$ c.m.
frame to the forward or backward hemisphere. The forward top
polarization is then defined as
\begin{eqnarray}
\emph{P}_t^{F}=\frac{\widehat{\sigma}_{-}(\cos\theta>0)-\widehat{\sigma}_{+}(\cos\theta>0)}
{\widehat{\sigma}_{-}(\cos\theta>0)+\widehat{\sigma}_{+}(\cos\theta>0)},
\end{eqnarray}
and the backward top polarization $\emph{P}_t^B$ is similarly
defined with $\cos\theta<0$. Here $\theta$ is the top scattering
angle in the $t\bar{t}$ c.m. frame.

Also, we will study the effects on the forward-backward asymmetry of the top
quark \cite{29}
\begin{eqnarray}
\emph{A}^t_{FB}=\frac{\widehat{\sigma}(\cos\theta>0)-\widehat{\sigma}(\cos\theta<0)}
{\widehat{\sigma}(\cos\theta>0)+\widehat{\sigma}(\cos\theta<0)},
\end{eqnarray}
which can be defined for either a fixed top quark helicity or with
the helicities summed, and $\theta$ is the top scattering angle in
the $t\bar{t}$ c.m. frame.
\begin{figure}[ht]
\begin{tabular}{cc}
\epsfig{file=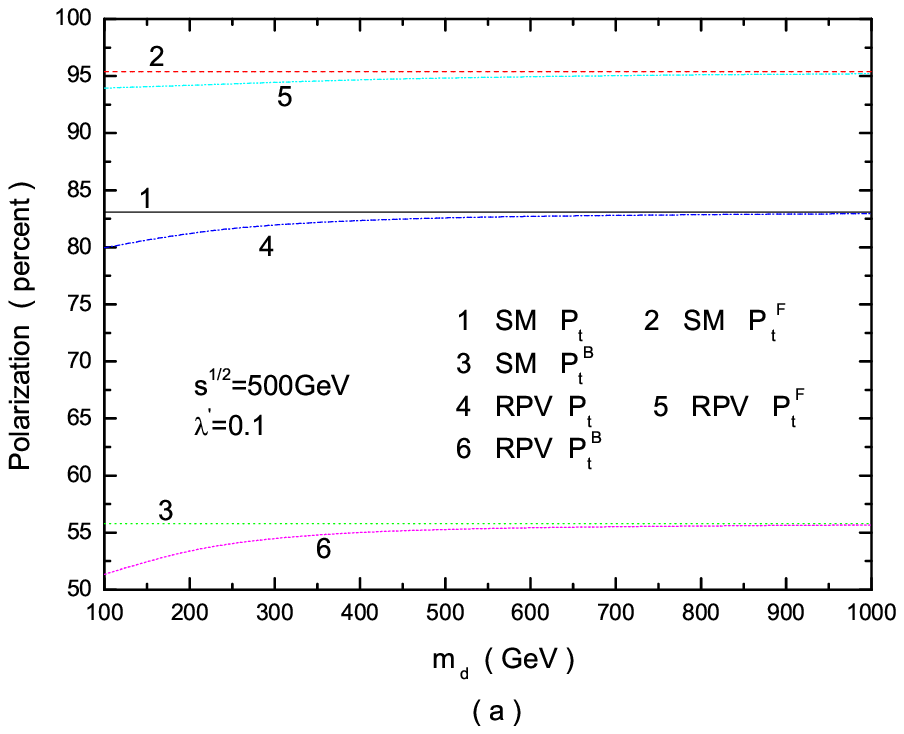,width=225pt,height=235pt}
\end{tabular}
\begin{tabular}{cc}
\epsfig{file=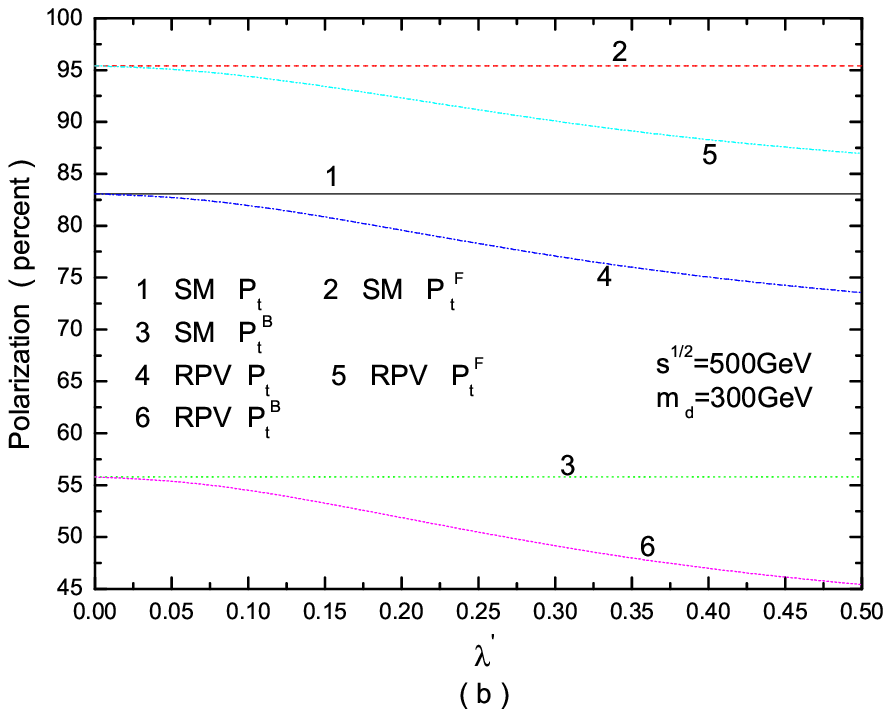,width=225pt,height=235pt}
\end{tabular}
\vspace*{-0.8cm} \caption{\small Top quark polarization versus (a)
the intermediate squark mass and (b) the \emph{R}-violating coupling
$\lambda^{'}$}
\end{figure} \label{fig4}
\begin{figure}[ht]
\begin{tabular}{cc}
\epsfig{file=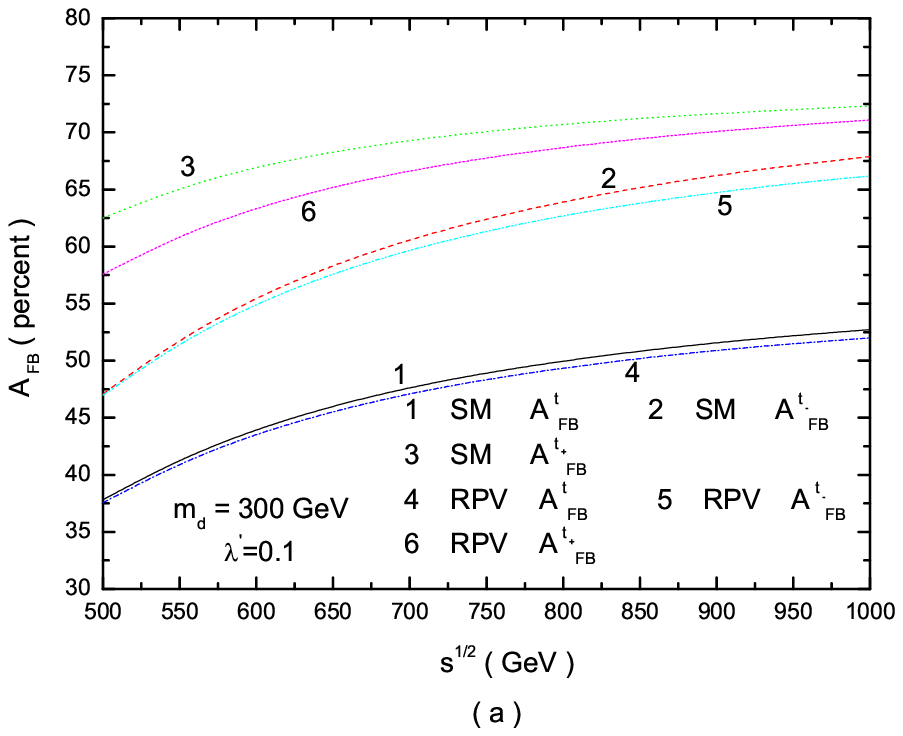,width=225pt,height=250pt}\\
\end{tabular}
\begin{tabular}{cc}
\epsfig{file=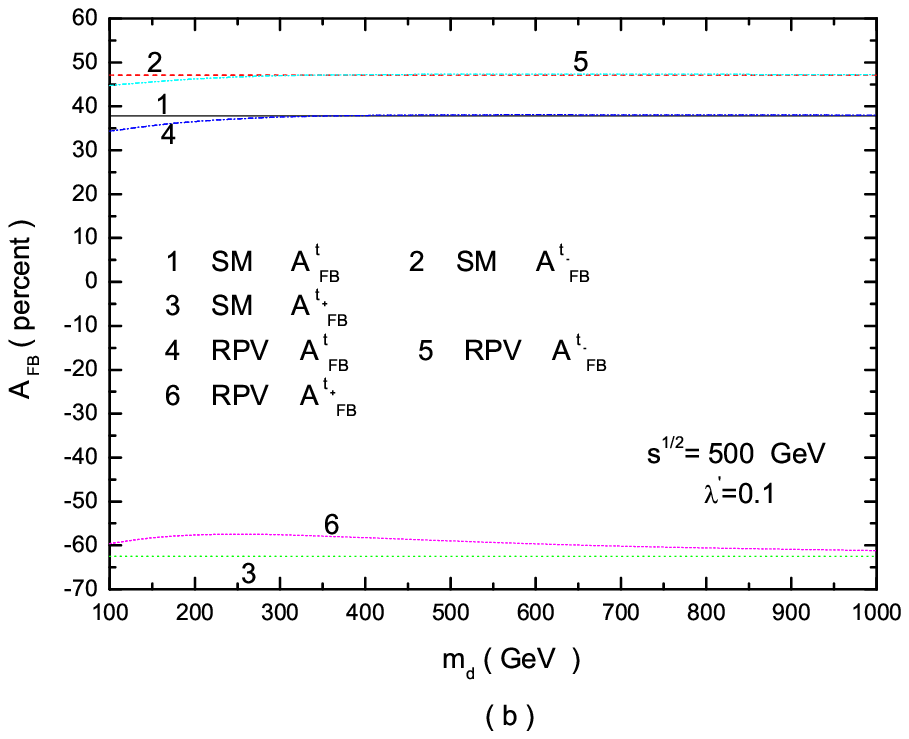,width=225pt,height=250pt}\\
\end{tabular}
\vspace*{-0.8cm}
\caption{\small Forward-backward asymmetries versus (a) the center-of-mass
energy and (b) the intermediate squark mass.} \label{fig5}
\end{figure}
In Fig.4(a), we plot the polarization versus the intermediate squark
mass. We see that as the squark mass increases, the polarizations
approach to their corresponding SM values. In Fig.4(b) we show the
polarization versus the \emph{R}-violating coupling. As the
\emph{R}-violating coupling increases, the polarizations become more
deviate from their SM values.

Fig.5(a,b) are the plots of forward-backward asymmetries versus the
center-of-mass energy and the squark mass. It is imperative to
mention that in Fig.5(a) we plotted the variation of the absolute
value of $A_{FB}$ with the center-of-mass energy. It is clear that
the absolute value of forward-backward asymmetries increase with the
center-of-mass energy.

The 2$\sigma$  statistical limit for $\lambda^{'}_{13k}$ versus
squark mass is shown in Fig.6 for the integrated luminosity of 100
$fb^{-1}$ with unpolarized and with left-handed polarized $e^{-}$
beams. We see that the linear collider can be quite powerful in
constraining these \emph{R}-violating couplings in case of
unobservation and the polarized beams are more powerful in probing
such \emph{R}-violating couplings.
\begin{figure}[ht]
\begin{center}
\epsfig{file=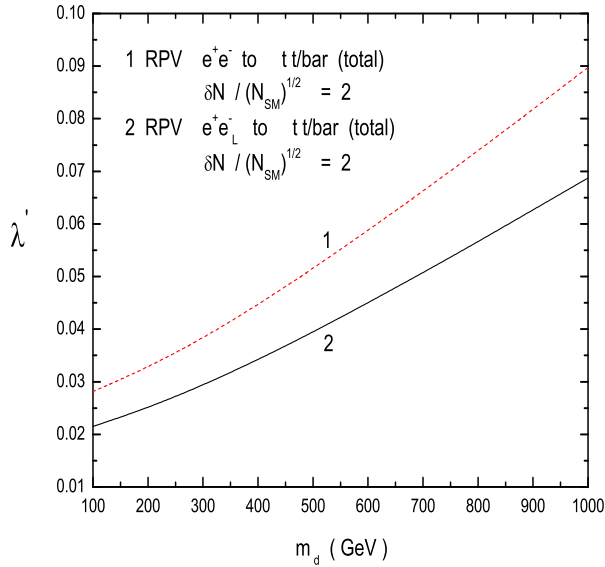,width=250pt,height=250pt}
\vspace*{-0.8cm}
\caption{\small
The $2\sigma$ statistical limit for $\lambda^{'}_{13k}$ versus squark mass
with center-of-mass energy of 500 GeV and a luminosity of 100 $fb^{-1}$.}
\label{fig6}
\end{center}
\end{figure}
\begin{center}
{\bf IV. Conclusion}
\end{center}

We studied the effects of the \emph{R}-parity violating interactions
in top pair production at a linear collider. We found that in the
allowed range of these \emph{R}-violating couplings, the top pair
production rate as well as the top quark polarization and the
forward-backward asymmetry can be significantly altered. By
comparing the unpolarized beams with the polarized beams, we found
that the polarized beams are more powerful in probing such new
physics.

\newpage
\begin{center}
{\bf Acknowlegement}
\end{center}

This work is supported by the National Natural Science
Foundation of China, the Excellent Youth Foundation of Henan
Scientific Committee and the Henan Innovation Project for University
Prominent Research Talents.

\end{document}